\documentclass[pra,aps,twocolumn,superscriptaddress]{revtex4-1}
\usepackage{amssymb,graphicx,color,bm,soul}
\usepackage{amsmath}
\usepackage{xcolor}

\newcommand{\bfv}{\boldsymbol}
\begin{document}
\title[The optical control of phase locking of polariton condensates]{The optical control of phase locking of polariton condensates}
\author{I. Y. Chestnov}
\email{igor_chestnov@westlake.edu.cn}
\affiliation{Westlake University, 18 Shilongshan Road, Hangzhou 310024, Zhejiang Province, China}
\affiliation{Institute of Natural Sciences, Westlake Institute for Advanced Study, 18 Shilongshan Road, Hangzhou 310024, Zhejiang Province, China}
\affiliation{Vladimir State University, Gorki St. 87, 600000, Vladimir, Russia}
\author{A. V. Kavokin}
\affiliation{Westlake University, 18 Shilongshan Road, Hangzhou 310024, Zhejiang Province, China}
\affiliation{Institute of Natural Sciences, Westlake Institute for Advanced Study, 18 Shilongshan Road, Hangzhou 310024, Zhejiang Province, China}
\author{A. V. Yulin}
\affiliation{National Research University of Information Technologies, Mechanics and Optics (ITMO University), Saint-Petersburg 197101, Russia}

\begin{abstract}
The phase and the frequency of an  exciton polariton condensate excited by a  nonresonant pump can be efficiently manipulated by an external coherent light. Being tuned close to the resonance with the condensate eigenfrequency, the external laser light imposes its frequency to the condensate and locks its phase, thereby manifesting a synchronization effect. The conditions of formation of the phase synchronized regime are determined. The synchronization of a couple of closely spaced polariton condensates by a spatially uniform coherent light is examined. At the moderate strength of the coherent driving the synchronization is accompanied by the appearance of symmetry-breaking states of the polariton dyad, while these states are superseded by the symmetric state at the high-intensity driving. By employing a zero-dimensional model of coupled dissipative oscillators with both dissipative and conservative coupling, we study the bifurcation scenario of the symmetry-breaking state formation.
\end{abstract}

\maketitle

\section{Introduction}

A strong light-matter interaction in semiconductor microcavities gives rise to formation of composite quasiparticles called exciton polaritons \cite{weisbuch1992}. Being a superposition of   excitons and   photons, they provide a strong optical non-linearity and may be characterized by a very light effective mass. These fascinating properties advantage polaritons over cold atoms in demonstrating collective many-body phenomena at high temperatures \cite{christopoulos2007,su2018}. However, the main difference of polaritonic systems from the conventional atomic condensates is a strong dissipation stemming from the finite lifetime of microcavity photons which necessitates external pumping to maintain polariton population.

The experimental methods for  excitation of coherent polaritons may be divided into two classes: non-resonant and resonant (or quasi-resonant). The first class may be also referred to as incoherent pumping where the phase of the forming condensate does not depend on the phase of the pump. This regime is frequently realized in polariton lasers \cite{christopoulos2007} that can be pumped both optically and electrically \cite{schneider2013}. The nonresonant electrical or optical pumping creates a  reservoir of incoherent polaritons. If the pumping strength exceeds the threshold value, the polaritons condense to a single quantum state \cite{kasprzak2006,sun2017}. The relaxation process is accelerated due to the bosonic stimulation by an occupancy of the final state.   Such a spontaneous buildup of a quantum coherence  is not phase-selective in the sense that the phase of the condensate is chosen spontaneously during the condensation.

When the condensation simultaneously occurs in several closely spaced condensation centers pinned to the system inhomogeneities, the interaction between condensates leads to  their mutual synchronization \cite{baas2008,wouters2008}. Although the total ensemble of the interacting condensates remains invariant to the global phase shift, the phase difference between neighbouring condensates is locked by the coupling \cite{ohadi2016}. Besides the synchronization can also take place between different polarization components of the spinor polariton condensate in the presence of intrinsic Josephson or spin-orbit coupling (caused by the TE-TM splitting) between the polarizations \cite{ohadi2015}.

{It is crucial that the coupling between the driven-dissipative condensates is inherently complex, i.e. it affects not only the  energy of the coupled state as the conventional conservative (Josephson) coupling typical for atomic condensates, but also the net losses \cite{aleiner2012}}. Since {  for  the driven dissipative systems the coupling  is determined self-consistently with the amplitudes of the condensates~\cite{kalinin2018networks}}, it is expected that several coupled condensates may be phase locked in various configurations which are characterized by different eigenfrequencies and condensation thresholds. The particular state of the ensemble   is chosen during the condensation according to the selection mechanism, which favors the state with the lowest polariton lasing threshold to grow faster than the other \cite{aleiner2012}. Recently it was demonstrated that under certain conditions the steady-state configuration of an ensemble of coupled polariton condensates can be associated with the global minimum of a particular spin Hamiltonian \cite{berloff2017,lagoudakis2017} assuming that the phases of the condensates are mapped to two-dimensional classical   spins. The phenomenon of establishment of a mutual coherent state of several polariton condensates can be considered as a synchronization of interacting polariton lasers by analogy with the coherent dynamics of arrays of interacting lasers \cite{Nixon2012,Gaeta2018}.


In contrast to the nonresonant pump, the quasiresonant excitation of polaritons with a coherent light provides a reliable tool of control of their properties. Namely, polaritons are formed in the state which assumes the frequency from the pump and is phase-locked to the pump. 
In this paper we address the problem of synchronization of the coherent polariton state created by nonresonant pumping to the coherent light having frequency that is close to the frequency of the condensate. In the simplest case of a single condensate whose eigenfrequency matches the  frequency of the laser light, the solution of this problem is trivial: the coherent excitation cancels the invariance to the global phase shift, thus the condensate is phase locked with the laser light \cite{kalinin2018Ising,caputo2019}. Here we demonstrate that even in the absence of a precise resonance, a coherent laser light is capable to  impose its frequency and its phase on the condensate. Drawing an analogy with the synchronization of a dissipative oscillator by a continuous driving force \cite{pikovsky2003}, we consider this problem in terms of the synchronization phenomenon.

The problem under study is relevant to the recent studies of phase-locked polariton condensates aimed at the realization of  polariton simulators. The manipulation of the  condensate phase by the coherent light can be associated with the action of an effective magnetic field on the particular pseudo-spin. The study of the interaction of the coherent light with an ensemble of coupled condensates is important as a tool of control over phase locking in an $XY$-polariton simulator. Recently the particular case of a coherent pumping at the frequency being a multiple to the condensate eigenfrequency was considered \cite{kalinin2018Ising}. {Here we consider different regimes of the synchronization of the coupled polariton condensates to the external near-resonant coherent drive. Special attention is paid to the  symmetry breaking bifurcation and the formation of the synchronized asymmetric states. }

The paper is organized as follows. Section \ref{SecII} presents the model system we consider. Section \ref{SecIII} describes the synchronization of a single nonresonantly excited condensate by the coherent pumping and determine the conditions of synchronization. In section \ref{SecIV} we extent the problem into the pair of coupled polariton condensates uniformly illuminated by the coherent light.
 Concluding the paper we discuss possible implementations of the predicted phenomenon.

\section{The system under study}\label{SecII}

We consider an exciton-polariton condensate  excited by the incoherent pump in a planar semiconductor microcavity in the presence of the low-intensity quasi-resonant coherent laser light.
The formation of the polariton condensate is described by the widely accepted mean-field model characterizing the polariton system by the complex wave function $\Psi$ of the coherent state and by the density $N_r$ of reservoir excitons \cite{wouters2007}:
\begin{subequations}\label{MainModel}
\begin{eqnarray}
i\hbar \partial_T \Psi &=& \left(- \frac{\hbar^2}{2m} \left(\partial_{XX} + \partial_{YY}\right) + g_c |\Psi|^2 + g_r N_r \right. \label{MainModela} \\
\nonumber &  & + \left. \frac{i\hbar}{2}\left(RN_r - \gamma_c\right) \right)\Psi + Fe^{-i\Delta t}, \\
\partial_T N_r &=& P  - \left(\gamma_r + R |\Psi|^2\right) N_r.\label{MainModelb}
\end{eqnarray}
\end{subequations}
Here $T$ is time, $m$ is the polariton effective mass, the terms  $g_c$ and $g_r$ are responsible for the energy shifts due to the polariton-polariton repulsion and the interaction with the reservoir, respectively. Populations of both condensate and reservoir dissipate with the rates $\gamma_c$ and $\gamma_r$, respectively. The net dissipation is balanced by the pump $P$  creating the incoherent excitons which then scatter to the condensate with the rate $RN_r|\Psi|^2$. The last term in the right-hand side of \eqref{MainModela} accounts for the spatially uniform coherent driving whose frequency is detuned from the bottom of the lower polariton branch by $\Delta$.

For the sake of simplicity of the following calculations we rewrite the equations \eqref{MainModela} and \eqref{MainModelb} in a dimensionless form:
\begin{subequations}\label{MainModelNorm}
\begin{eqnarray}
i\partial_t \psi &=& \left(  -\frac{1}{2} \left(\partial_{xx} + \partial_{yy}\right) + i(n_r-1)   + \right. \\
\notag && \left. \vphantom{-\frac{1}{2}} + |\psi|^2 + g n_r \right) \psi  + fe^{-i\delta t}, \label{NormEq1} \\
\partial_t n_r &=& p -\left( \gamma + \beta |\psi|^2 \right)n_r, \label{NormEq2}
\end{eqnarray}
\end{subequations}
where we have redefined $t=T\gamma_c/2$, $x=X\sqrt{m\gamma_c\left/2\hbar \right.}$,
$y=Y\sqrt{m\gamma_c\left/2\hbar \right.}$,
$\psi=\Psi\sqrt{{2g_c}\left/{\hbar\gamma_c}\right.}$,
$n_r= RN_r/\gamma_c$,
$g=2g_r\left/\hbar R \right.$,
$\gamma=2\gamma_r\left/\gamma_c\right.$,
$\beta=\hbar R\left/ g_c\right.$,
$p=2RP\left/\gamma_c^2\right.$,
$f=F\sqrt{{8g_c}\left/{\hbar\gamma_c^3}\right.} $ and
$\delta=2\Delta\left/\gamma_c \right.$.

\section{Synchronization of a single polariton condensate by a coherent laser light}\label{SecIII}
We start considering the interaction of a single polariton condensate with the external coherent light.  We assume that the  nonresonant pump has a Gaussian shape $p(\mathbf{r})=p_0 \exp\left( -\frac{(\mathbf{r} - \mathbf{r}_0)^2}{2l^2} \right)$, where $\mathbf{r}$ is an in-plane vector. In the absence of the resonant excitation the condensate is formed in the steady state  $\psi(\mathbf{r},t) = \psi(\mathbf{r})\exp(-i\mu t)$ provided that the incoherent pump amplitude  $p_0$ exceeds the threshold value $p_{\rm th}$. 
In the framework of the mean-field model \eqref{MainModelNorm}, the spontaneous choice of the condensate phase  which can be parameterized as $\varphi = \rm{arg}\left(\psi(\mathbf{r}=\mathbf{r}_0)\right)$   is  provided by tacking a white noise distribution of the polariton field at the initial moment of time.

\begin{figure}
  \includegraphics[width=\linewidth]{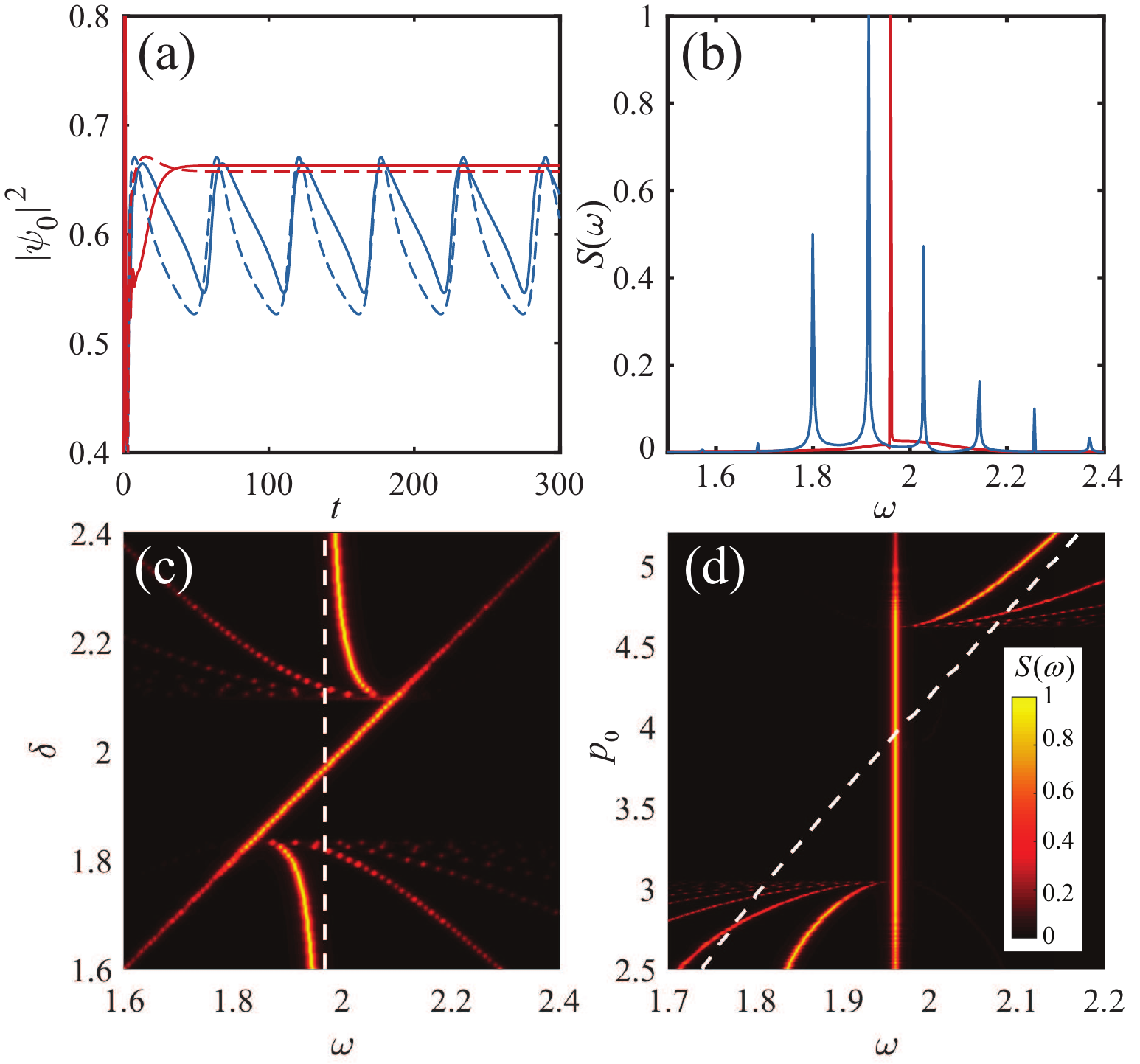}
  \caption{Synchronization of a single polariton condensate by an external coherent light. Panels (a) and (b) show the dynamics of the condensate density maximum $|\psi_0|^2=|\psi(\mathbf{r}=\mathbf{r}_0)|^2$ and the corresponding temporal spectra $S(\omega)=\int \psi\left(\mathbf{r}=\mathbf{r}_0\right) e^{-i \omega t} dt$ normalized to unity. The red curves correspond to the driving frequency $\delta=1.96$ which is close to the condensate eigenfrequency at $p_0=4$, while the blue curves correspond to the strongly nonresonant driving $\delta=1.8$. 
The solid curves were calculated with use of the full model \eqref{MainModelNorm}. The dashed curves correspond  to the predictions of the simplified model \eqref{Eq_toy_1}. (c) The evolution of the condensate spectral density $S(\omega)$ as the driving frequency  $\delta$ varies while amplitude of the nonresonant   pump  is fixed to $p_0=4$. (d) The same as on (c) but when the amplitude of the incoherent pump $p_0$ varies while $\delta=1.96$. The white dashed lines on (c) and (d) indicate  the condensate eigenfrequency $\mu$. For all panels $g=0.46$, $\beta=2.5$, $\gamma=0.4$, $f=0.07$ and $l=0.93$.}\label{Fig:SingleCondSynch}
\end{figure}

As the coherent laser pumping is switched on, it tends to impose its frequency and phase on polaritons. However, the weak driving is unable to impose its phase to the condensate,  provided that the frequency mismatch between the laser and the  condensate  is large. Instead, it perturbs the condensate steady state inducing oscillations of polariton density, -- see the blue curve in Fig.~\ref{Fig:SingleCondSynch}(a). These oscillations can be detected by the frequency comb in the temporal spectrum of the  condensate (see Fig.~\ref{Fig:SingleCondSynch}(b)). The oscillations are governed by the nonlinear mixing of the condensate eigenfrequency $\mu$ and the  driving frequency $\delta$.

As the driving frequency  approaches the frequency of the unperturbed condensate, the steady state  restores, see the red curve in figure \ref{Fig:SingleCondSynch}(a). The condensate frequency merges with the frequency of the coherent pump, $\mu=\delta$, manifesting itself the synchronization between the condensate and the light, see the red line in Fig.~\ref{Fig:SingleCondSynch}(b). Note that the merging of the spectral lines happens regardless of whether the coherent pump is tuned below or above the condensate frequency. Indeed, the range of the driving frequencies providing synchronization is shown in
Fig.~\ref{Fig:SingleCondSynch}(c), which illustrates the evolution of the condensate spectrum as the driving frequency $\delta$ scans from the red-detuned to the blue-detuned region, while the driving amplitude $f$  remains fixed. The spectral width of the  synchronization domain is dependent on $f$ as well as on the  properties of the condensate. For instance, the variation of the amplitude of the incoherent pump $p_0$,  that corresponds to the variation of the condensate eigenfrequency $\mu$, also reveals  synchronization in the finite range of pumping amplitudes, see Fig.~\ref{Fig:SingleCondSynch}(d).

To achieve analytical results and to explain the synchronization phenomenon in simple terms we simplify the model \eqref{MainModelNorm} by substituting it with a single ordinary differential equation for the complex amplitude $A$ of the polariton field:
\begin{eqnarray}\label{Eq_toy_1}
i\partial_t A = \left( \alpha |A|^2 -\tilde \delta  +i\Gamma - i\nu |A|^2   \right) A +\tilde f,
\end{eqnarray}
which was written in the reference frame rotating with the effective driving frequency $\tilde\delta$. Here we have introduced the  net gain $\Gamma$, the saturation of the gain $\nu|A|^2$ accounting for the effect of   the reservoir depletion,  the effective nonlinear frequency shift $\alpha$ and  the effective driving force $\tilde f$. The correspondence between the parameters of the model \eqref{MainModelNorm} and their counterparts in \eqref{Eq_toy_1} is discussed in Appendix A.

Equation \eqref{Eq_toy_1} is relatively easy to analyze. In the absence of the coherent pump, the steady state solution reads: $A=\sqrt{n_0}\exp\left({-i\mu_0 t}\right)$, where  $n_0=\Gamma/\nu$ and $\mu_0 =\alpha n_0 -  \tilde\delta$ in the rotating frame corresponds to the frequency mismatch between the driving force and the condensate eigenfrequency. In the presence of the coherent pumping the only possible steady state is characterized by the eigenfrequency equal to the driving frequency, $\mu_0=0$. Thus seeking the solution in the form $A(t)=\sqrt{n}\exp\left[i\varphi(t)\right]$ and assuming weakness of the coherent pump in the sense that it does not affect the condensate amplitude, $n=n_0$, we get
\begin{equation}
\sqrt{n_0}\partial_t \varphi = \left(\tilde \delta - \alpha n_0 \right) \sqrt{n_0} - \tilde f\cos\left( \varphi \right).
\end{equation}
The stable stationary solution of this equation, $\partial_t \varphi = 0$, corresponds to the establishment of the synchronization regime. It requires
\begin{equation}\label{sync_condition_simple}
\tilde f \geq \left|\tilde \delta- \alpha\Gamma/\nu \right|\sqrt{\Gamma/\nu}.
\end{equation}
This simple synchronization condition defines the critical value of the  driving strength necessary  to slave polariton condensate.
It implies that in the case of the weak driving the synchronization occurs either in the vicinity of the resonance, $\tilde \delta = \alpha\Gamma/\nu$, or close to the condensation threshold where the condensate occupancy $\Gamma/\nu$ is low. In the close proximity of the threshold, $\Gamma/\nu \rightarrow0$, the condensate can not resist synchronization  even if the  driving is weak and strongly detuned. Equation \eqref{sync_condition_simple} indicates also that the synchronization condition is insensitive to the sign of the frequency mismatch.

A strict analysis of the synchronized (steady-state) solution  of equation \eqref{Eq_toy_1} yields  the following equation for the condensate density:
\begin{equation}\label{SingleCondAmplitude}
\left(\alpha n - \tilde \delta\right)^2 n +  \left(\Gamma-\nu n \right)^2n = {\tilde f}^2,
\end{equation}
which may have {either one or} three real roots. Let us note that this approach accounts for the influence of the coherent pump on the amplitude of the condensate and thus is more general than the approach based on the assumption that only the phase but not the density of the condensate is affected by the driving force.

\begin{figure}
\includegraphics[width=\linewidth]{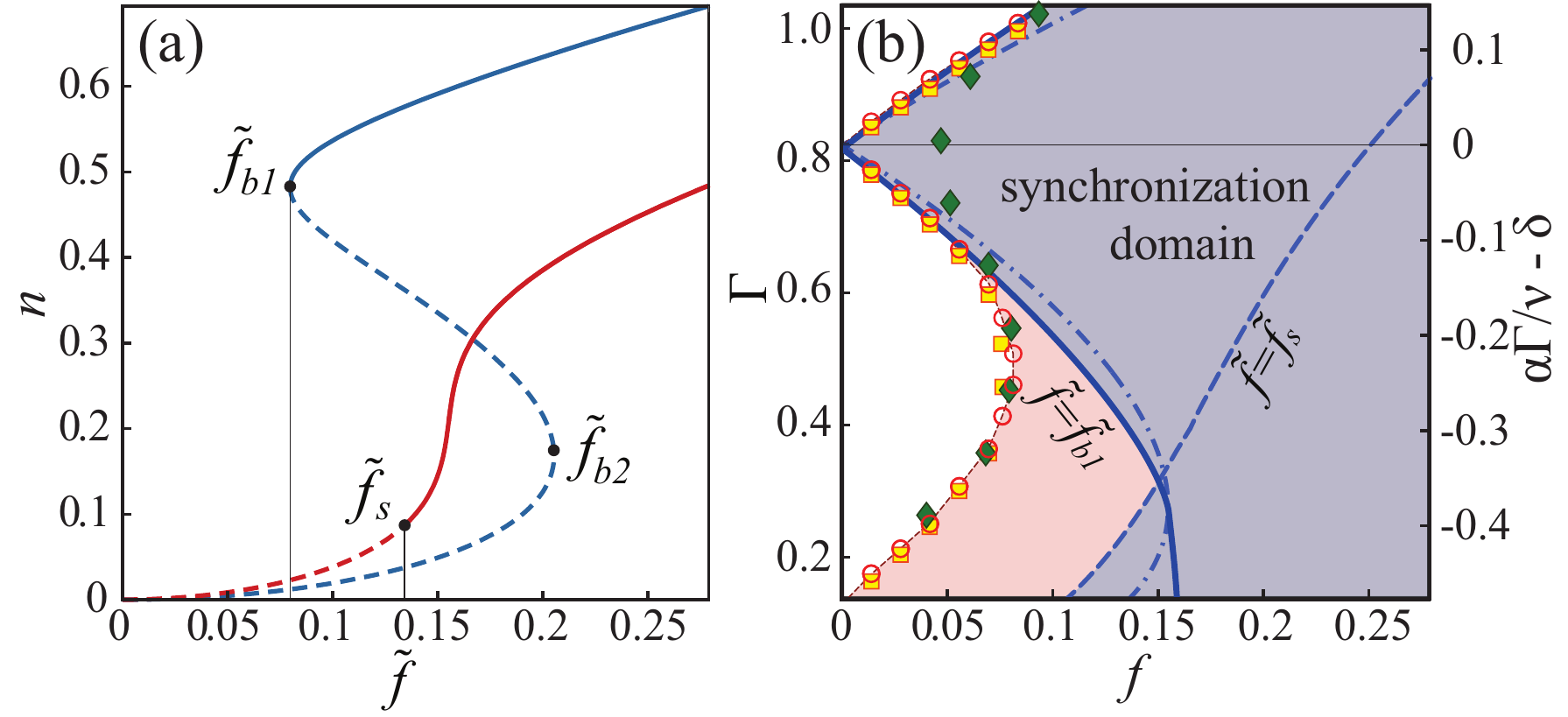}
\caption{(a) The $n(\tilde f)$-dependence of the steady-state solution of  \eqref{Eq_toy_1}. The stable states corresponding to the synchronization with the coherent light are shown with solid curves while the dashed curves indicate unstable solutions. The gain parameter is $\Gamma=0.6$ for the blue curve and $\Gamma=0.24$ for the red curve. (b) The phase diagram for the synchronization of a single polariton condensate with the coherent light. The purple region corresponds to the synchronization domain predicted by the toy model \eqref{Eq_toy_1}. The red region framed by the red circles is obtained  from the numerical solution of the full system \eqref{MainModelNorm}. The simplified condition \eqref{sync_condition_simple} is shown with the dash-dotted curve. The yellow squares show the boundary of the synchronization domain under the homogeneous illumination of the symmetric polariton dyad (the interspot distance is $d=17$, see Fig.~\ref{Fig:TwoCondSynch}(a)), while the green diamonds indicate the same for the antisymmetric dyad ($d=14$).  For all data $\delta=1.96$. The rest parameters for both panels are the same as in Fig.~\ref{Fig:SingleCondSynch}.} \label{Fig:SynchDomain}
\end{figure}

The typical shape of the $n(\tilde f)$-dependence for the solution is shown in Fig.~\ref{Fig:SynchDomain}(a). {Although the steady-state solutions exist for any value of the coherent pumping strength, the synchronous regime is established only for those states which are stable against small perturbations.} The analysis of the Lyapunov exponents  yields that {the synchronized solution} is stable, provided that {
\begin{subequations}
\begin{eqnarray}\label{StabConditions}
&2\nu n > \Gamma, \label{StabCondition1}\\
&3\left(\alpha^2+\nu^2\right)n^2-4\left(\Gamma\nu+\alpha\delta\right)n +\delta^2+\Gamma^2 >0.\label{StabCondition2}
\end{eqnarray}
\end{subequations}
}%
\noindent {These conditions are simultaneously satisfied for the upper branch of the S-shaped curve (see the blue curve in Fig.~\ref{Fig:SynchDomain}(a)). However } at the left folding point $\tilde f=\tilde f_{b2}$  the solution becomes unstable and the synchronized state transforms into a state with two incommensurable frequencies. Mathematically it means that the pair of stable and unstable fixed points corresponding to the upper and the intermediate branches of the S-shaped curve collide at $\tilde f = \tilde f_b$ and disappear { giving birth to a stable limit cycle}. The birth of the limit cycle manifests itself in the spectral frequency comb shown in Fig.~\ref{Fig:SingleCondSynch}(b).

Note that the lower branch of the S-shaped curve does not support synchronization {for arbitrary small values of the driving strength} as it is typically unstable in a stark contract with the conventional optical bistability regime \cite{baas2004}, that is realized in the absence of the incoherent pumping. {In particular, the whole lower branch is unstable provided that $\Gamma>2\nu n_{b2}$ (see the condition \eqref{StabCondition1}), where $n_{b2}$ determines the condensate density at the right bending point, see  Fig.~\ref{Fig:SynchDomain}(a). In this case,} the position of the left bending point $\tilde f_{b1}$, which can be easily obtained from \eqref{SingleCondAmplitude} taking $\partial_n \tilde f=0$, should be considered as the critical value of the coherent pump amplitude above which the synchronization occurs.

{If equation \eqref{SingleCondAmplitude} admits a single real solution (see the red curve in Fig.~\ref{Fig:SynchDomain}(a)), the inequality \eqref{StabCondition2} always holds and the stability of the synchronized state is governed by the condition
\eqref{StabCondition1}. }
Thus, the solution is also unstable at the weak driving strength, when the steady state condensate population $n$  vanishes. However the condensate  becomes stable {through a supercritical Hopf bifurcation} at $\tilde f \geq \tilde f_s$ manifesting establishment of the synchronization regime.  The critical driving strength $ \tilde f_s \equiv \tilde f(n_s)$  is determined from \eqref{SingleCondAmplitude} taking
$n=n_s=\Gamma/2\nu$.

The inequalities $\tilde f>\tilde f_s(\tilde \delta)$ and $\tilde f > \tilde f_{b2}(\tilde \delta)$ constitute the synchronization conditions which are shown in the phase diagram Fig.~\ref{Fig:SynchDomain}(b) plotted on the parameter plane $ (\Gamma, f )$. Note that the simplified condition \eqref{sync_condition_simple} describes the synchronization phenomenon well enough, especially in the vicinity of the resonance, $\alpha\Gamma/\nu = \tilde\delta$, see the dash-dotted curve.

For the sake of comparison, the results of numerical simulations of the full 2D  model \eqref{MainModelNorm} are also shown in the phase diagram, Fig.~\ref{Fig:SynchDomain}, see the red circles indicating the boundary of the synchronization domain. The best coincidence between the models is achieved in the vicinity of the resonance. However the region close to the condensation threshold is worse described by the toy model \eqref{Eq_toy_1} since it factors out the reservoir-induced blue shift. The blue-shift is accounted for in the effective detuning $\tilde \delta$ (see Appendix A).
Far from the threshold this assumption is justified by the effect of  depletion of the reservoir.  Indeed, if $\beta |\psi|^2 \gg \gamma$, the reservoir density $n_r=p/\left(\gamma+\beta\left|\psi\right|^2\right)$ is depleted. In contrast, if the condensate density $|\psi|^2$  is comparable to or lower than $\gamma/\beta $, the impact of the reservoir-induced blue shift is crucial.  As a result,  the simplified model overestimates the value of the frequency mismatch $ \alpha\Gamma/\nu - \tilde \delta$ near the threshold shifting the boundary of the synchronization domain  towards the higher values of the driving strength, see Fig.~\ref{Fig:SynchDomain}(b).

\section{Polariton dyad  in the presence of a coherent laser light}\label{SecIV}
\subsection{Synchronization of the symmetric and antisymmetric polariton dyad to the coherent light}

\begin{figure}
\includegraphics[width=\linewidth]{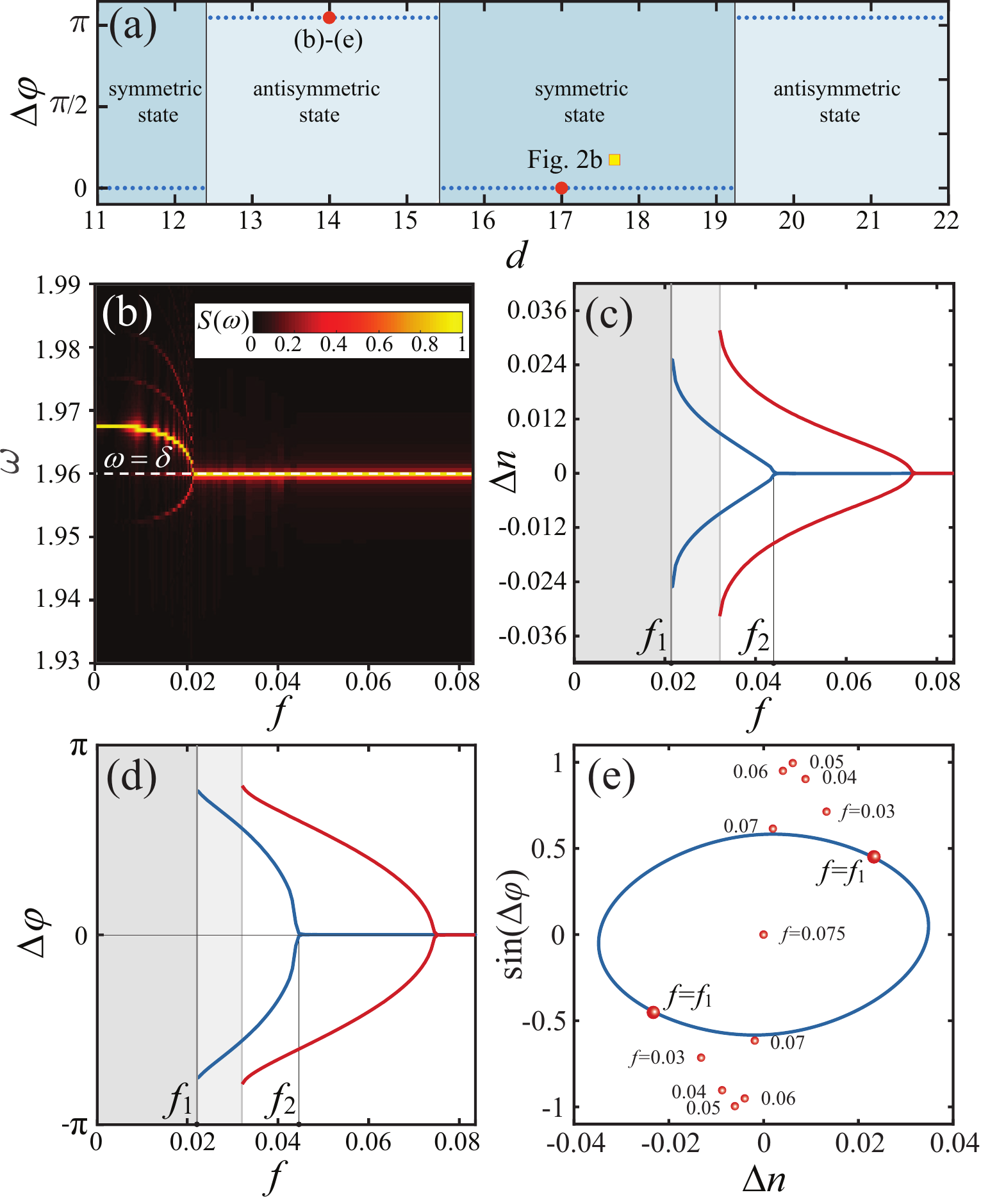}
\caption{Synchronization of the polariton dyad with the coherent laser light. (a) Switching between the symmetric and antisymmetric states of the dyad with the growth of the inter-condensate distance $d$ {for the fixed incoherent pump power}. (b) The spectral density $S(\omega)$ calculated at $\mathbf{r}=\mathbf{r}_2$ from the numerical simulations of the dynamics of the full model. Panels (c) and (d) show the dependencies of the steady state population imbalance $\Delta n$ and the phase difference $\Delta \varphi$ on the coherent pump amplitude $f$, respectively. The blue curves correspond  to the numerical solution of the full model \eqref{MainModelNorm} while the red curves show the predictions of the equations \eqref{Eq_toy_2_1} and \eqref{Eq_toy_2_2}. (e) The blue line shows the limit cycle trajectory on the $\left(\Delta n, \sin(\Delta \varphi)\right)$ plane in the close proximity of the bifurcation point $f=f_1$. Large red dots correspond to the steady-state solutions exactly at the bifurcation point. Stable symmetry broken states, which appear at $f\in \left[f_1,f_2\right)$,  are shown with small dots. The parameters are: $p_0=4$, $\delta=1.96$, $g=0.46$, $\beta=2.5$, $\gamma=0.4$ and $l_1=l_2=0.93$. } \label{Fig:TwoCondSynch}
\end{figure}

Another fundamental question is what happens when  more than one condensate is illuminated by a coherent laser light.
We proceed considering the simplest case of a polariton dyad created by two spatially separated spots of the incoherent pump. In particular, we  take  $p(\mathbf{r})=p_1 \exp\left( -\frac{(\mathbf{r} - \mathbf{r}_1)^2}{2l_1^2} \right) + p_2 \exp\left( -\frac{(\mathbf{r} - \mathbf{r}_2)^2}{2l_2^2} \right)$, with $d=\left|\mathbf{r}_1 -\mathbf{r}_2 \right|$ being the interspot distance. Because of the  outflow of polaritons from under the pump spots, the condensates interact with each other. In the absence of the coherent pump this coupling causes the formation of a mutually synchronized (coherent) state. If the pumps are identical ($p_1=p_2$ and $l_1=l_2$),  polaritons may condense either in in-phase (symmetric) or in anti-phase (antisymmetric) configurations depending on the  mutual coupling strength, which is governed by the interspot  distance $d$ and  the velocity of the polariton outflow \cite{ohadi2016}. In particular, the symmetric and antisymmetric states alternate  as the interspot distance varies while the nonresonant pump power remains fixed, see Fig.~\ref{Fig:TwoCondSynch}(a).  Note  that in both configurations the condensates are equally populated. For the sake of simplicity  we leave out of the scope of this paper the class of symmetry broken solutions which appear in the weak lasing regime \cite{aleiner2012}.

The external {coherent driving tends to lock phases of the condensates.} Hence it may alter the structure of the dyad. In this paper we focus on the case of a uniform illumination assuming that the coherent light excites the microcavity at the normal incidence. Experimentally this is feasible with strongly defocused laser beam.

{We start analyzing the synchronization scenario in the framework of the full model \eqref{MainModelNorm}.} It is anticipated, that, by analogy with the case of a single condensate, a  weak driving is unable to synchronize the dyad. At the same time,  an intense coherent driving should dominate over the inter-condensate coupling suppressing the intrinsic structure of the dyad and synchronizing both condensates. 
However, the behaviour at the intermediate driving strength lacks of understanding. Nevertheless, it is obvious that the  synchronization scenario should depend on whether the dyad is symmetric or antisymmetric in the absence of the coherent light.

When the nonresonant pump  provides the in-phase configuration of the polariton dyad, the synchronization scenario is pretty much the same as for a single condensate although the synchronization {conditions} are slightly affected by the inter-condensate coupling (see the yellow squares in Fig.~\ref{Fig:SynchDomain}(b)). {Indeed, in the absence of the coherent pump the coupling  alters the frequency of the dyad and the amplitudes of the condensates in respect to the case of a stand-alone polariton condensate. Since the dyad state remains symmetric for any value of the driving amplitude (the condensates are in-phase and equally populated both in synchronous and asynchronous regimes), the symmetric dyad can be associated to a single driven oscillator which effective frequency and the loss rate are affected by the complex inter-condensate coupling.}

The case of  the condensates  phase locked  with  $\pi$-phase difference in the absence of the coherent light is much richer. {The typical synchronization scenario for this case is illustrated in Fig.~\ref{Fig:TwoCondSynch}. The weak driving induces oscillations of the condensate amplitude. The increase of the driving strength  results in synchronization of the dyad to the coherent pump.} From symmetry reasons, it is obvious that the antisymmetric state  cannot be synchronized by the homogeneous driving without modification of the entire structure of the solution.  Instead, it is anticipated that  the {a pair of identical} condensates {illuminated by the homogeneous coherent pumping} should be synchronized in the symmetric state.

However, it appears that, as the  coherent driving strength increases above the critical value $f_1$ (see Fig.~\ref{Fig:TwoCondSynch}), the limit cycle, which corresponds to the {oscillating} regime, is superseded by the  steady state { with \textit{broken symmetry}}. Despite of the fact that the incoherent pump spots are identical, the condensates have different populations and their relative phase equals neither  $0$  or $\pi$ (see Fig.~\ref{Fig:TwoCondSynch}(c) and Fig.~\ref{Fig:TwoCondSynch}(d)).

With the further increase of the coherent pump amplitude, the degree of asymmetry of the state, i.e. the population imbalance $\Delta n=n_1-n_2$ and the phase difference  $\Delta \varphi = \varphi_1 - \varphi_2$ decrease. Here $n_{1,2}\equiv n\left(\mathbf{r}=\mathbf{r}_{1,2}\right)$ and $\varphi_{1,2}\equiv \varphi\left(\mathbf{r}=\mathbf{r}_{1,2}\right)$. Both $\Delta n$ and $\Delta\varphi$ vanish at $f=f_2$ as a second bifurcation happens and the symmetric steady state forms, see figures \ref{Fig:TwoCondSynch}(c) and \ref{Fig:TwoCondSynch}(d). In this regime, the coherent pump dominates over inter-condensate interactions and governs the phase configuration of the dyad. 

\subsection{Synchronization scenario in terms of the coupled oscillators model}

To study the synchronization scenario  in more details we again resort to the simplified model considering the interacting condensates as a pair of linearly coupled oscillators:
\begin{subequations}\label{Eq_toy_2}
\begin{eqnarray}
&&i\partial_t A_1 = \left(\alpha |A_1|^2 -\tilde \delta   +i\Gamma   - i\nu |A|^2    \right) A_1  + \sigma A_2  +\tilde f, \label{Eq_toy_2_1} \\
&&i\partial_t A_2 = \left(\alpha |A_2|^2-\tilde \delta   +i\Gamma   - i\nu |A|^2   \right) A_2   +\sigma A_1 +\tilde f. \label{Eq_toy_2_2}
\end{eqnarray}
\end{subequations}
We assume that the coupling parameter is complex, $\sigma =  \sigma_{\rm J}+i\sigma_{\rm d} $. The conservative (Josephson) coupling $\sigma_{\rm J}$  results in the frequency splitting between the symmetric and antisymmetric states. The $\sigma_{\rm d}$ component is responsible for the dissipative coupling which accounts for the fact that the net losses of the coupled state  do depend on the relative phase. The numerical values of these parameters are obtained by fitting of the results of the simulations of the full 2D model \eqref{MainModelNorm} to the results predicted by the ordinary differential equations \eqref{Eq_toy_2_1} and \eqref{Eq_toy_2_2}. The details of this fitting procedure are presented in Appendix A. 

{In what follows we focus on the case of the antisymmetric polariton dyad which corresponds to the positive dissipative coupling parameter, $\sigma_{\rm d}>0$.} The synchronization scenario for this case, predicted by model \eqref{Eq_toy_2_1} and \eqref{Eq_toy_2_2}, is shown in Fig.~\ref{Fig:TwoCondSynch} with red lines {while the predictions of the full model \eqref{MainModelNorm} correspond to the blue curves}. Although the simplified model fails to predict the position of the bifurcation points $f_{1,2}$ accurately, both the spontaneous symmetry breaking phenomenon and the break of the synchronous regime are reproduced qualitatively correctly. Thus we believe that the analysis of the simplified model is capable to describe the behavior of the full system. {The discrepancy between the models should be attributed to the renormalization of the coupling  parameter $\sigma$ by the presence of the homogeneous driving. Indeed, the coupling strength is determined by the interference of the overlapping condensate wave functions. An intensive homogeneous resonant pumping modifies the distribution of the condensate phase affecting thereby the overlap between neighbouring condensates and altering the coupling parameter.} %

The spontaneous symmetry breaking scenario, which is realized as the pump intensity decreases below $f=f_2$, indicates the presence of a supercritical pitchfork bifurcation. {This statement can be proved in terms of the coupled oscillators model \eqref{Eq_toy_2_1} and \eqref{Eq_toy_2_2} applying} a perturbation theory in the vicinity of the bifurcation point $f_2$. The details of the calculations are accumulated in  Appendix B.
In the leading  order of approximation the deviation $\Delta A_{1, \, 2}$ of the symmetry broken solution $(A_1, A_2)$ from the symmetric one   can be found in the form $\Delta A_{1, \, 2}=a \left( \xi_{1,3} + i \xi_{2,4} \right) $ where $\xi_{1, 2,3,4}$ are real constants defined in Appendix B.
The equation for the amplitude $a$ has the form
\begin{equation}
\partial_t a=\lambda a + \epsilon a^3 \label{norm_form}
\end{equation}
with real coefficients $\lambda$ and $\epsilon$.  
Equation \eqref{norm_form} is the normal form of a pitchfork bifurcation. 
Calculation of the coefficients entering   \eqref{norm_form} yields $\epsilon <0$. So, the symmetry breaking indeed goes through supercritical pitchfork bifurcation.

{The analysis of the dynamics predicted by the coupled oscillators model reveals that} the bifurcation  switching the system between the synchronous and asynchronous (oscillating) regimes  at $f=f_1$ is an infinite-period bifurcation, which consists in the appearance of the  two pairs of stable and unstable states on the limit cycle. It destroys the motion around the limit cycle trajectory and the system switches to the one of the two stable fixed points  corresponding to the solutions with broken symmetry. The limit cycle trajectory in the vicinity of the bifurcation point calculated with the coupled oscillators model  is shown in Fig.~\ref{Fig:TwoCondSynch}(e). The large red dots on the phase trajectory mark the position the stable fixed points at $ f=f_1$.

{Note that the described synchronization scenario is robust against the change of the domain of the inter-condensate distances $d$ which provide the antisymmetric configuration of the undriven dyad, see Fig.~\ref{Fig:TwoCondSynch}(a). However the threshold value of the coherent pump intensity corresponding to the establishment of the synchronous regime, $f=f_1$, is sensitive to the parameters of the system, see the green diamonds in Fig.~\ref{Fig:SynchDomain}. In particular, it depends on the frequency mismatch between the coherent pumping frequency and the condensate and on the inter-condensate distance $d$ which governs the inter-condensate coupling}. Besides, in the case of a strong frequency mismatch, the spontaneous symmetry breaking bifurcation and the bifurcation accompanying formation of the synchronous state merge. {In this case the symmetry broken state disappears and the oscillating regime, typical for the weak coherent pumping, is directly superseded by the symmetric state, $\delta \varphi=0$, as the driving strength increases above $f_2$. In order to determine the conditions of existence of the symmetry broken state we further simplify the coupled oscillators model reducing it to the model of symmetrically driven coupled phase oscillators. }

\subsection{Coupled phase oscillators model}
By analogy with the case of a single oscillator (see equation \eqref{sync_condition_simple}), we assume that the driving strength is weak and does not affect the condensate populations, $\left|A_1\right|=\left|A_2\right|=\sqrt{\Gamma/\nu}$. Thus, taking $A_{1,2}=\left|A_{1,2}\right| \exp\left[{i\varphi_{1,2}}\right]$, the state of the dyad is described by the coupled equations for the condensate phases:
\begin{eqnarray}\label{Eq_phase_osc}
\partial_t\varphi_{1,2} &=& q - \sigma_{\rm J} \cos\left(\varphi_{2,1} - \varphi_{1,2}\right) + \\
\nonumber &+& \sigma_{\rm d} \sin\left(\varphi_{2,1} - \varphi_{1,2}\right) -  {f}^\prime\cos \varphi_{1,2},
\end{eqnarray}
where $q= \tilde{\delta} - \alpha \Gamma/\nu$ is an effective detuning and $f^\prime=\tilde{f}\sqrt{\nu/\Gamma} $. In the absence of the driving,  $f^\prime=0$, these equations predict the formation the symmetric state $\varphi_2-\varphi_1=0$ at positive dissipative coupling $\sigma_{\rm d}>0$ and the antisymmetric state  $\varphi_2-\varphi_1=\pi$ at $\sigma_{\rm d}<0$.

\begin{figure}
\includegraphics[width=\linewidth]{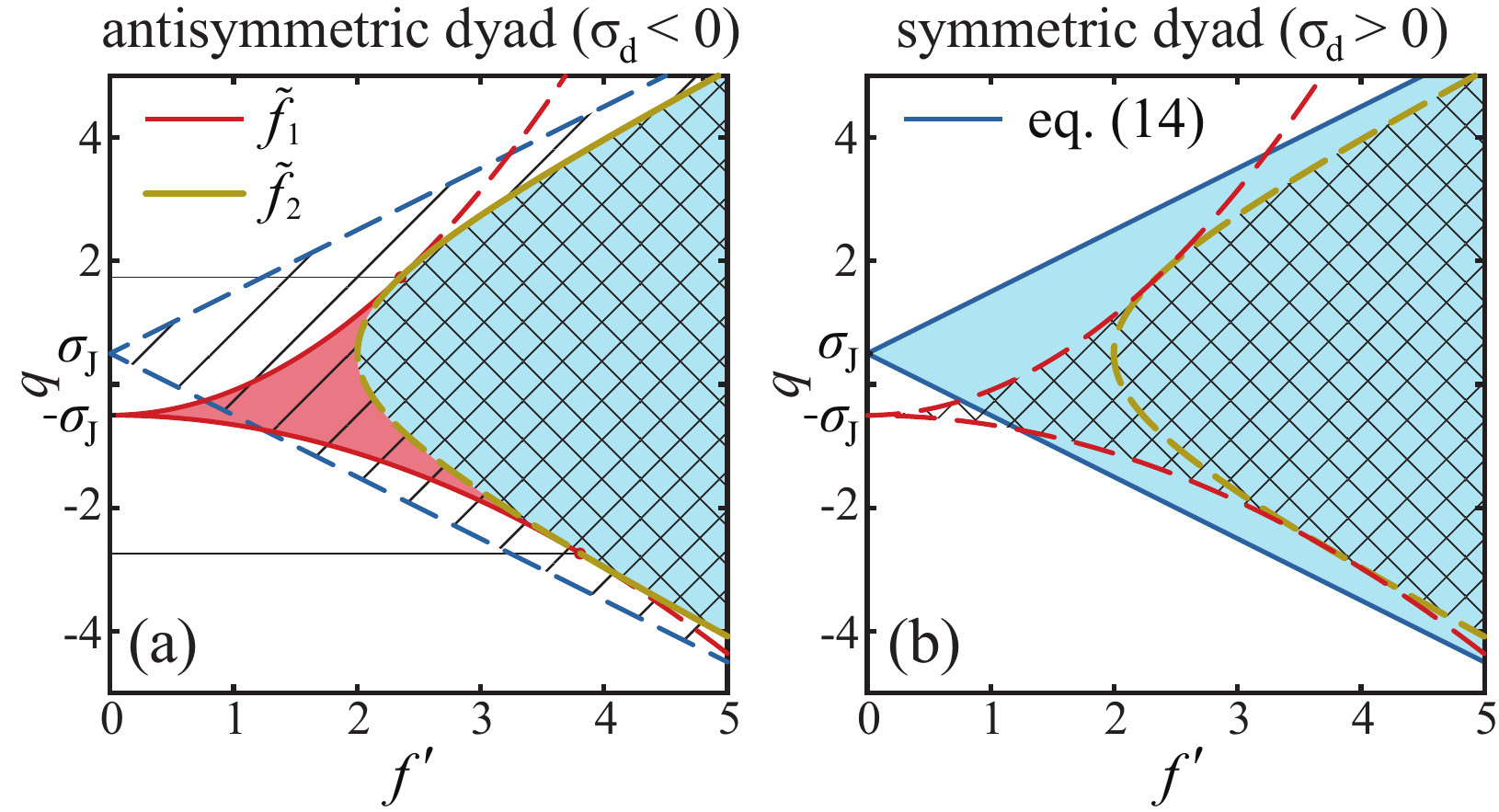}
\caption{Bifurcation diagram of the stationary solutions of the coupled phase oscillators model \eqref{Eq_phase_osc}. (a) For $\sigma_{\rm d} =- 1$ and  $\sigma_{\rm J} = 0.5$. (b) For $\sigma_{\rm d} =  1$ and  $\sigma_{\rm J} = 0.5$. The shaded  domains correspond to the stable states: the blue one to the symmetric state and the red one to the symmetry broken states. The red line corresponds to condition \eqref{eq_bs_driving}, the yellow one -- to \eqref{eq_f2}, the black horizontal lines  -- to condition \eqref{eq_bs_detuning} and the blue one -- to the boundary given by \eqref{eq_symstate_cond}. The solid segments of the curves indicate the threshold driving required for synchronization. The hatched domains correspond to dynamically unstable states: the diagonal hatch to the unstable symmetric state and the square hatch to the unstable symmetry broken state. } \label{Fig:PhaseModelDiagram}
\end{figure}

Being quite simple, model \eqref{Eq_phase_osc} predicts the same synchronization scenario as observed in 2D case. In particular, equations \eqref{Eq_phase_osc} have both symmetric $\varphi_1=\varphi_2$ and the symmetry broken $\varphi_1 \neq \varphi_2$ stationary solutions, see  Appendix C. The corresponding bifurcation diagram is shown in Fig.~\ref{Fig:PhaseModelDiagram}.

The analysis shows that the solution with broken symmetry appears provided that the coherent driving strength exceeds the threshold value, which is defined as 
\begin{equation}\label{eq_bs_driving}
\tilde{ f}_1 = 2 \sqrt{\Gamma/\nu}\left(\sqrt{ \left( \sigma_{\rm J}+q \right)^2 \left( \sigma_{\rm J}^2+\sigma_{\rm d}^2 \right)} - \sigma_{\rm J}\left(\sigma_{\rm J}+q\right) \right)^{1/2}
\end{equation}
for the detuning $q$ belonging to the range
\begin{equation}\label{eq_bs_detuning}
\left|q+\sigma_{\rm J}\right| < 2\sqrt{\sigma_{\rm J}^2 + \sigma_{\rm d}^2}.
\end{equation}

Although the stationary solutions with  broken symmetry exist at any value of the complex coupling, for the  dyad which is symmetric at zero driving (i.e. at $\sigma_{\rm d}>0$) they are always dynamically unstable and hence these synchronous states cannot be observed experimentally. In contrast, the antisymmetric dyad ($\sigma_{\rm d}<0$) {can transform into the stable states with broken symmetry.} Besides, it can be synchronized in the symmetry broken state even at the vanishing driving strength provided that its frequency matches the eigenfrequency of the stand-alone polariton dyad, namely, at $q=-\sigma_{\rm J}$, see \eqref{eq_bs_driving} and Fig.~\ref{Fig:PhaseModelDiagram}a. In the latter case the symmetry broken state bifurcates from the antisymmetric  solution $\varphi_2-\varphi_1=\pi$ at $f^{\prime}=0$. With the increase of the driving strength the degree of asymmetry of the state, i.e. the phase difference  $\varphi_2-\varphi_1$, decreases until the symmetry broken states disappear at the pitchfork bifurcation (which is of supercritical type at $\sigma_{\rm d}<0$ and of subcritical type in the opposite case) at $f^\prime = f^\prime_2$, where:
\begin{equation}\label{eq_f2}
\tilde{f}_2=\sqrt{\Gamma/\nu}\sqrt{4\sigma_{\rm d}^2 + \left(q-\sigma_{\rm J}\right)^2}.
\end{equation}

Note that the stable symmetry broken solutions do not appear at any driving strength if condition \eqref{eq_bs_detuning} violates, i.e. if the laser frequency is strongly detuned from the  eigenfrequency of the stand-alone antisymmetric dyad. {However other symmetry broken solutions exist at $f^\prime>f^\prime_2$ for arbitrary detuning, but these solutions are always dynamically unstable (see Appendix C for the details)}.

Besides the symmetry broken solutions there are also the symmetric states which appear provided that
\begin{equation}\label{eq_symstate_cond}
\tilde{f}>\left| q-\sigma_{\rm J}\right| \sqrt{\Gamma/\nu}.
\end{equation}
At $\sigma_{\rm d}>0$ a pair of symmetric solutions is always stable. Both these states have the same frequency inherited from the coherent driving, although they are characterized by the different phase shift in respect to the phase of the laser light.  In this case condition \eqref{eq_symstate_cond} determines the synchronization threshold and is equivalent to \eqref{sync_condition_simple}.

In the case of the  dyad which is antisymmetric at zero driving ($\sigma_{\rm d}<0$), the symmetric synchronized states become stable only at $f^\prime>f^\prime_2$, see the blue shaded region in Fig.~\ref{Fig:PhaseModelDiagram}. Thus for the strongly non-resonant  driving the synchronization of the antisymmetric dyad requires much stronger power than in the case of the symmetric one.

The domains of existence of the synchronized solutions defined by the conditions \eqref{eq_bs_driving}-\eqref{eq_symstate_cond} remain invariant to the change of a sign of the dissipative coupling $\sigma_{\rm d}$. However, the stability properties of the solutions are different for the symmetric and the antisymmetric dyad.  Besides,  the change of a sign of the Josephson coupling $\sigma_{\rm J}$ leads to the reflection of the bifurcation diagram about $q=0$ axis. Note that both dissipative and Josephson coupling parameters varies with the variation of the inter-condensate distance and the pumping power, see e.g.~\cite{ohadi2016,kalinin2018matter}.

\section{Conclusion and outlook}

We describe a novel mechanism of manipulation of non-equilibrium condensates of exciton polaritons formed in semiconductor microcavities. This mechanism, based upon a very general nonlinear effect of synchronization with an external coherent driving, paves the way to the non-resonant control of the properties of driven-dissipative polariton condensates.  In particular, in the synchronization regime the frequency of the polariton condensate does not depend on the intensity of the incoherent pump but is equal to the coherent pump frequency. In this regime the phase of the polariton condensate is locked to the phase of the coherent pump. Thus, this effect allows to control the phase of the condensate by detuned external laser beam. A simple model describing the spatially distributed condensate by its amplitude can qualitatively reproduce the observed phenomena.

The interplay of mutual synchronization of a couple of neighbouring condensates and their synchronization to the external coherent light is also studied. Those condensates  which are  locked in-phase in the absence of the coherent driving, can be easily synchronized to the coherent light. However, the synchronization of the antisymmetric polariton dyad is accompanied by the appearance of the symmetry broken configuration of neighbouring polariton condensates which mediates the formation of a mutual coherent state. This intermediate regime is characterized by the imbalance of the condensate populations and corresponds to the  spontaneous breaking of the state symmetry. It is superseded by the symmetric (in-phase) configuration of the polariton dyad as the intensity of the coherent light grows. {The obtained results can be straightforwardly generalized to the case of an ensemble of coupled polariton condensates using the model of driven coupled dissipative oscillators, which allows reducing the problem to the simplified dynamical model.}

In spite of  the fact that the properties of the polariton dyad are dependent of the spatial overlap of the  wave functions of the condensates, the model of two coupled nonlinear oscillators driven by an external force are capable to describe all the peculiarities of the synchronization scenario.  Thus the model of coupled oscillators represents an effective tool for investigation of the polariton dynamics.
%
%
In particular, it assists the analysis of the bifurcations happening in the polariton systems. In our case the model of coupled oscillators allows to reduce  the spontaneous  symmetry breaking bifurcation { to the normal form of a supercritical pitchfork bifurcation.}  

Synchronization of a single condensate by the coherent light may assist an experimental measurement of the relative phases between the condensates in the  ensemble created by the multi-spot nonresonant pumping. Fixing the phase of a single condensate one may use the quasi-resonant laser light as a reference beam.

Besides, it is worth mentioning that the effect of synchronization can potentially be used for manipulation of the spin-polarized polariton condensates which are created by the polarized nonresonant laser pumping. { In particular, it has been recently demonstrated \cite{sakaguchi2017} that the spin-up and spin-down components having different eigenfrequencies can be mutually synchronized due to spin-orbit interaction. The approach developed in the present paper can be used for the studies of how these vector polariton  states can be synchronized to coherent pumps of different polarizations and frequencies.}


We also believe that the effect of synchronization can assist to control the distribution of phases and amplitudes in the arrays of polariton condensates which can be useful, for example, for the engineering of polariton lasing systems consisting of polariton laser arrays and for polariton simulators. In particular, since the external coherent pump  affects the condensate phase, it can be associated with the effective magnetic field acting either on a single effective spin or on the several condensates in the array. 

\begin{acknowledgments}
This work is supported by Westlake University (Project No. 041020100118). AVY was financially supported by the Government of the Russian Federation (Grant 074-U01) through ITMO  Fellowship schemes. IYC acknowledges the support from RFBR, grants No.  17-52-10006, 17-42-330928 and the Ministry of Education and Science of the Russian Federation, Project No. 16.1123.2017/4.6.
\end{acknowledgments}

\appendix

\section{Derivation of the simplified zero-dimensional model of synchronization }

\begin{figure*}
\includegraphics[width=0.75\linewidth]{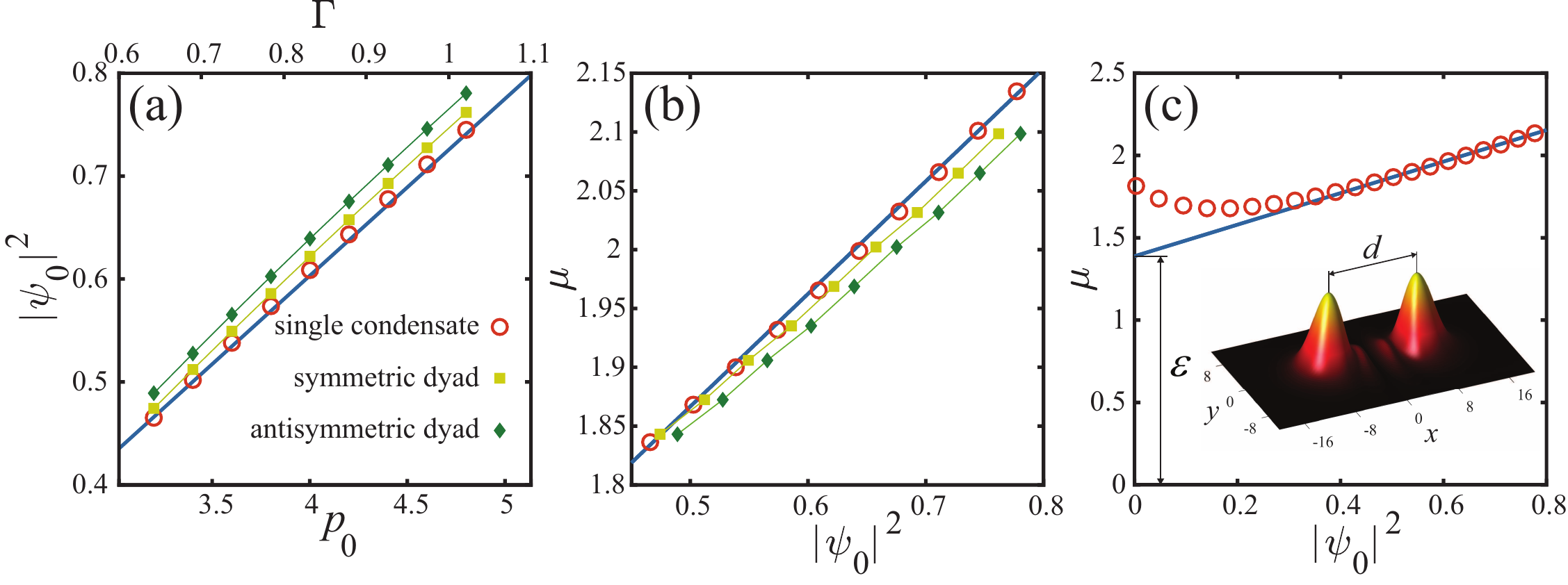}
  \caption{(a) Dependence of the condensate density maximum $|\psi_0|^2$ on the  amplitude $p_0$ of the incoherent pump (bottom scale).  The blue line shows the $\left|A\left(\Gamma\right)\right|^2$-dependence predicted by the single oscillator model \eqref{Eq_toy_1}. The corresponding values of the effective gain $\Gamma$ are shown on the top scale.   Red circles correspond to the properties of a single stand-alone condensate predicted by the partial differential equations model \eqref{MainModelNorm}.  Yellow squares (symmetric state) and green diamonds (antisymmetric state) correspond to the steady-state configurations of the polariton dyad in the absence of coherent pump. (b) Dependence of the frequency $\mu$ on the density maximum $|\psi_0|^2$. All markers have the same meaning as in  panel (a). (c) The same as in panel (b) but for a wider range of $|\psi_0|^2$. The shift of the condensate eigenfrequency  $\varepsilon$ at the threshold ($|\psi_0|^2=0$) is chosen in such a way that $\mu (\left|A\right|^2 )$-dependence (blue line) calculated with the model \eqref{Eq_toy_1} fits the predictions of the full model (circles) far above the threshold. The discrepancy between the models in the vicinity of the threshold should be attributed to the depletion of the reservoir. The inset illustrates the typical shape of the polariton density distribution, $|\psi_0|^2$. }
  \label{Fig:Fitting}
\end{figure*}

Although the toy models \eqref{Eq_toy_1} and \eqref{Eq_toy_2_1},\eqref{Eq_toy_2_2} are capable to explain all the peculiarities of the synchronization phenomenon, the quantitative agreement with the full partial differential equations \eqref{MainModelNorm} may be achieved  tacking the correct values of the system parameters. In general, these parameters can be determined  integrating out spatial degrees of freedom. Instead of approximating the condensate wave function and calculating overlap integrals \cite{kalinin2018networks}, we resort to fitting of  parameters of the simplified models using the predictions of the full model as a reference. The good agreement between the predictions of the simplified and full models shown in Fig.~\ref{Fig:SynchDomain}(b) justifies our approach.

The  polariton field amplitude $A$ entering the simplified models is associated with the maximum of polariton density. Namely, $A=\psi_0=\psi(\mathbf{r}=\mathbf{r}_0)$ for the single oscillator model   and $A_{1,2}=\psi(\mathbf{r}=\mathbf{r}_{1,2})$ for the coupled oscillators model.

For the model  \eqref{Eq_toy_1} the effective linear gain $\Gamma$ is related to the amplitude of the incoherent pump by the relation $\Gamma= b_1 p_0- b_2$, where $b_1$ and $b_2$  can be found from the growth rate of the polariton density at the linear stage of the polariton formation and from the position of the polariton formation threshold, respectively. For the considered parameters (see the caption of Fig.~\ref{Fig:SingleCondSynch}) we obtain $b_1=0.237$ and $b_2=0.116$.
The nonlinear saturation of the gain $\nu$ and the nonlinear frequency shift $\alpha$ are obtained fitting the dependencies of the condensate density maximum $|\psi_0|^2$ on the pump amplitude $p_0$ [Fig.~\ref{Fig:Fitting}(a)] and of the condensate eigenfrequency $\mu$  on the  density maximum $|\psi_0|^2$ [Fig.~\ref{Fig:Fitting}(b)], respectively.  The best fit is achieved for $\alpha=0.957$ and $\nu=1.378$. The effective driving force $\tilde f$ is related to the amplitude of the coherent pump  as $\tilde f = k f$, where the factor $k$ is obtained from the fit of the width of the synchronization domain, Fig.~\ref{Fig:SynchDomain}(b). For the considered parameters we get $k=1.1$.

Note that the  model \eqref{Eq_toy_1} excludes the excitonic reservoir from the consideration. However, the reservoir is responsible for the blue shift of the condensate eigenfrequency, which may be significant even at the condensation threshold. At the same time, the synchronization condition is crucially affected by the detuning of the coherent light frequency from the condensate eigenfrequency. Thus the reservoir-induced frequency shift has to be taken into account in model \eqref{Eq_toy_1}. For this reason we use the effective frequency detuning $\tilde \delta=\delta-\varepsilon$, where $\varepsilon$  corresponds to the  eigenfrequency shift at the condensation threshold, see Fig.~\ref{Fig:Fitting}(c). For the considered parameters $\varepsilon=1.388$.

 Describing synchronization of the polariton dyad, we assume that the parameters of the model \eqref{Eq_toy_2_1} and \eqref{Eq_toy_2_2} are the same as for the case of a single condensate. This assumption is justified by the weak spatial overlap between the condensate wave functions for the inter-condensate distances considered here, see the inset to Fig.~\ref{Fig:Fitting}(c). The mutual coupling between spatially separated condensates, which is accounted  by $\sigma_{\rm J}$ and $\sigma_{\rm d}$ parameters,  affects the population and the frequency of the condensate. In particular, the steady state solutions of equations \eqref{Eq_toy_2_1} and \eqref{Eq_toy_2_2} for the symmetric and antisymmetric configurations of the polariton dyad read $\left|A_{\rm s,as}\right|^2=\left|A_{0}\right|^2 \pm \sigma_{\rm d}\left/ \nu \right.$ and $\mu_{\rm s,as} = \mu_0 \pm \alpha \sigma_{\rm d} \left/\nu\right. \pm \sigma_{\rm J}$, where $\left|A_{0}\right|^2$ and $\mu_0$ characterize the state of a stand-alone condensate. With these solutions one can determine the dissipative $\sigma_{\rm d}$ and Josephson $\sigma_{\rm J}$ coupling parameters using the amplitude and the frequency calculated with the full model \eqref{MainModelNorm}, see figures \ref{Fig:Fitting}(a) and \ref{Fig:Fitting}(b). In particular, at $p_0=4$ we get $\sigma_{\rm J}=0.0292$ and $\sigma_{\rm d}=-0.042$.

At the end of this section, we report the dimensional parameters of the governing model \eqref{MainModela},\eqref{MainModelb}, which determines the dimensionless parameters of the normalized equations~\eqref{NormEq1},\eqref{NormEq2}: $\gamma_c=0.5$~${\rm ps}^{-1}$, $\gamma_r=0.1$~${\rm ps}^{-1}$, $R=$~$\mu{\rm m}^2{\rm ps}^{-1}$, $g_c=0.02$~${\rm meV}$, $g_r=0.0175$~${\rm meV}$.

\section{Normal form of the spontaneous symmetry breaking bifurcation}
\setcounter{section}{2}

In this section we present the perturbation approach which  reduces equations \eqref{Eq_toy_2_1} and \eqref{Eq_toy_2_2} to the normal form \eqref{norm_form} in the vicinity of the bifurcation point $\tilde f=\tilde f_2$. It appears convenient to reformulate the problem in the vector form. The polariton state is described by $\bfv{U}_i=\left( \Re A_1, \Im A_1, \Re A_2, \Im A_2 \right)^{T}$ and governed by the equation
\begin{eqnarray}
\partial_t  \bfv{U}_{i} =  \bfv{N}_{i} [ \vec U] + \tilde f  \bfv{ u}_i,
\label{appB1}
\end{eqnarray}
where $ \bfv{N}_i$ is a vector with components being functions of $\vec U$  and $\bfv{u}_i=(0, -1, 0, -1)^{T}$. The nonlinear operator $\bfv{N}_i$ is defined as
\begin{eqnarray} \bfv{N}_i[\vec X]= \hat M \vec X +
                       \left(\begin{array}{c}
                       (X_1^2+X_1^2)(\alpha X_2 - \nu X_1) \\
                       (X_1^2+X_1^2)(-\alpha X_1 - \nu X_2) \\
                       (X_3^2+X_4^2)(\alpha X_4 - \nu X_3) \\
                       (X_3^3+X_4^2)(-\alpha X_3 - \nu X_4) \\
                       \end{array}                   \right), \nonumber
\end{eqnarray}
where
\begin{eqnarray} \hat M = \left(
                     \begin{array}{cccc}
                       \Gamma & - \tilde\delta & \sigma_{\rm d} & \sigma_{\rm J} \\
                       \tilde\delta & \Gamma & - \sigma_{\rm J} & \sigma_{\rm d} \\
                       \sigma_{\rm d} & -\sigma_{\rm J} & \Gamma & - \tilde\delta \\
                       \sigma_{\rm J}  & \sigma_{\rm d} & \tilde\delta & \Gamma\\
                     \end{array}                   \right).   \nonumber
\end{eqnarray}

Here we denote the vector quantities by bold face letters with subscripts. For the arguments of vector functions we use the notation of vector as an arrowed letter meaning that each of the components of the vector function depend  on all components of the argument. Square brackets are used to denote the arguments of the vector functions.

A symmetric solution $A_1=A_2=A_{\rm s}$  of the equations \eqref{Eq_toy_2_1} and \eqref{Eq_toy_2_2}  is given by
\begin{eqnarray}
 \left(i \Gamma+ (\alpha - i\nu)  |A_{\rm s}|^2 -\tilde \delta  + \sigma \right) A_{\rm s}  = - \tilde f.
 \label{appB6}
\end{eqnarray}
%
To account for the dynamics in the vicinity of the bifurcation point we
 develop the perturbation theory looking for a solution in the form 
\begin{eqnarray}
\bfv{U}_i= \bfv{Y}_i +\bfv{y}_i,
\label{appB1_}
\end{eqnarray}
where $\bfv {Y}_i [\tilde f]=\left( \Re A_{\rm s}, \Im A_{\rm s}, \Re A_{\rm s}, \Im A_{\rm s} \right)^{T}$ describes the symmetric steady state  and $\bfv{y}_i$ is a small perturbations of the stationary solution. To describe the steady states bifurcating from $ \bfv{Y}_i$ the equation for the correction has to contain the terms nonlinear in respect to $\bfv{ y}_i$. We assume that the deviation from the bifurcation point $\Delta \tilde f = \tilde f-\tilde f_2$ is small and thus we keep only the terms linear in respect to $\Delta \tilde f$. Then by expansion in Taylor series we obtain
\begin{eqnarray}
&&\bfv{ N}_i[\vec Y[\tilde f]+\vec y] = 
\bfv{ N}_i [ \vec Y [\tilde f_2] +\vec y]] + \frac{\partial \bfv{ N}_i [\vec Y[\tilde f_2]]}{\partial \bfv{ Y}_j}   \frac{\bfv{ Y}_j[\tilde f_2]}{\partial \tilde f}\Delta f + \nonumber \\
&&+ \frac{\partial^2 \bfv{ N}_i [\vec Y[\tilde f_2]]}{\partial \bfv{ Y}_j \partial \bfv{Y}_k}   \frac{\bfv{ Y}_j[\tilde f_2]}{\partial \tilde f}\Delta \tilde f \bfv{y}_k.
\label{appB2}
\end{eqnarray}

Since $\vec Y $ is a stationary solution, we have $\bfv{N}_{i} [ \vec Y] = - \tilde f  \bfv{ u}_i$. 
This allows to transform (\ref{appB2}) to
\begin{eqnarray}
\bfv{ N}_i[\vec Y[f]+\vec y] =
\bfv{ N}_i [ \vec Y [\tilde f_2] +\vec y]] - \Delta \tilde{ f} \bfv{ u}_i  + \nonumber \\
+\frac{\partial^2 \bfv{ N}_i [\vec Y[\tilde f_2]]}{\partial \bfv{ Y}_j \partial \bfv{Y}_k}   \frac{\bfv{ Y}_j[\tilde f_2]}{\partial \tilde f}\Delta \tilde f \bfv{y}_k.
\label{appB3}
\end{eqnarray}

Substituting (\ref{appB1_}) and (\ref{appB3}) into (\ref{appB1}) we obtain
\begin{eqnarray}
\partial_t  \bfv{y}_{i} =  \bfv{ N}_i [ \vec Y [\tilde f_2] +\vec y] -\bfv{ N}_i [ \vec Y [\tilde f_2]]  + \nonumber \\
+\frac{\partial^2 \bfv{ N}_i [\vec Y[\tilde f_2]]}{\partial \bfv{ Y}_j \partial \bfv{Y}_k}   \frac{\bfv{ Y}_j[\tilde f_2]}{\partial \tilde f}\Delta \tilde f \bfv{y}_k.
\label{appB4}
\end{eqnarray}

The fact that the nonlinearity in the equation \eqref{appB4} is cubic makes it convenient to write it in the form
\begin{eqnarray}
\partial_t  \bfv{y}_{i} = N_{i, j , k, l}  (\tilde{\bfv{Y}}_{j}  +\bfv{y}_{j})(\tilde{\bfv{Y}}_{k}  +\bfv{y}_{k})(\tilde{\bfv{Y}}_{l}  +\bfv{y}_{l}) - \nonumber \\
- N_{i, j , k, l}  \tilde{\bfv{Y}}_{j}  \tilde{\bfv{Y}}_{k} \tilde{\bfv{Y}}_{l} +M_{i, j} \bfv{y}_{j}+\Delta \tilde f \Lambda_{i, j}  \bfv{y}_j,
\label{appB5}
\end{eqnarray}
with   $$\Lambda_{i, j}=\frac{\partial^2 \bfv{ N}_i [\vec Y[\tilde f_2]]}{\partial \bfv{ Y}_j \partial \bfv{Y}_k}   \frac{\bfv{ Y}_k[\tilde f_2]}{\partial \tilde f}$$ and $N_{i, j, k, l}=\tilde N_{i, j, k, l} + \bar  N_{i, j, k, l} $ accounts for the conservative part of the nonlinearity by $\tilde N_{i, j, k, l}$ and for the dissipative nonlinearity by  $\bar  N_{i, j, k, l} $. In the symmetric form the nonzero components of $\tilde N_{i, j, k, l}$ are
$\tilde N_{1, 2, 2, 2}=\tilde N_{3, 4, 4, 4}=\alpha$, $\tilde N_{2, 1, 1, 1}=\tilde N_{4, 3, 3, 3}=-\alpha$,
$\tilde N_{1, 1, 1, 2}=\tilde N_{1, 1, 2, 1}=\tilde N_{1, 2, 1, 1}= \tilde N_{3, 3, 3, 4}=\tilde N_{3, 3, 4, 3}=\tilde N_{3, 4, 3, 3}= \frac{1}{3}\alpha$,
$\tilde N_{2, 1, 2, 2}=\tilde N_{2, 2, 1, 2}=\tilde N_{2, 2, 2, 1}= \tilde N_{4, 3, 4, 4}=\tilde N_{4, 4, 3, 4}=\tilde N_{4, 4, 4, 3}= -\frac{1}{3}\alpha$.
For the nonzero components of $\bar N_{i, j, k, l}$ we have
$\bar N_{1, 1, 1, 1}=\bar N_{2, 2, 2, 2}=\bar N_{3, 3, 3, 3}=\bar N_{4, 4, 4, 4}=-\nu$,
$\bar N_{1, 1, 2, 2}=\bar N_{1, 2, 1, 2}=\bar N_{1, 2, 2, 1}=\bar N_{2, 1, 1, 2}=\bar N_{2, 1, 2, 1}=\bar N_{2, 2, 1, 1}=
\bar N_{3, 3, 4, 4}=\bar N_{3, 4, 3, 4}=\bar N_{3, 4, 4, 3}=\bar N_{4, 3, 3, 4}=\bar N_{4, 3, 4, 3}=\bar N_{4, 4, 3, 3}=
-\frac{1}{3}\nu$.

The operator $\hat L$ governing the stability of polariton state according to \eqref{appB5} is $\hat L= \hat M + 3\bfv{N}_{i, j, k, l}\bfv{Y}_k\bfv{Y}_l +\Delta \tilde f \Lambda_{i, j}$. At the point $\tilde f=\tilde f_2$ the operator $\hat L[\tilde f_2]=\hat L_0$ has one zero eigenvalue with an eigenvector $\xi$. We scale the eigenvectors so that their norm is equal to $1$. Then we look for the solution $\vec y$ in the form
$\vec y = a \vec \xi + \vec w$ where $\vec w$ is a vector orthogonal to $\vec \xi$, i.e. $(\vec \xi^{\dag}, \vec w)=0$, where the  dagger symbol denotes the eigenvectors of the adjoin operators and the parenthesis with comma denote the scalar product.
In this notations \eqref{appB5} takes the form
\begin{eqnarray}
&\vec \xi \partial_t  a + \partial_t  \vec w = a \Delta \tilde f \hat \Lambda \vec \xi + \Delta \tilde f \hat \Lambda \vec w + \hat L_0 \bfv{w}_{i} + 3 a^2 N_{i, j , k, l}  \bfv{Y}_{j} \bfv{\xi}_k \bfv{\xi}_l  \nonumber \\
&+ 6 a N_{i, j , k, l}  \bfv{Y}_{j} \bfv{\xi}_k \bfv{w}_l + 3 N_{i, j , k, l}  \bfv{Y}_{j} \bfv{w}_k \bfv{w}_l + a^3 N_{i, j , k, l}  \bfv{\xi}_{j} \bfv{w\xi}_k \bfv{\xi}_l  \nonumber \\
&+3 a^2 N_{i, j , k, l}  \bfv{\xi}_{j} \bfv{\xi}_k \bfv{w}_l + 3 a N_{i, j , k, l}  \bfv{\xi}_{j} \bfv{w}_k \bfv{w}_l 
+N_{i, j , k, l}  \bfv{w}_{j} \bfv{w}_k \bfv{w}_l. \nonumber \\
\vphantom{.}
\label{appB7}
\end{eqnarray}

In the first order approximation we obtain: 
$$ \partial_t  a  = \lambda a ,$$
where $\lambda[\tilde f]=\Delta \tilde  f  \left( \vec \xi^{\dag}, \hat \Lambda \vec \xi \right)$. In this case $\lambda$ corresponds to the first order approximation of eigenvalue generating the instability. Instead of calculation of the eigenvalue perturbatively, it can be calculated solving the corresponding spectral problem. The component $\vec w$ is equal to zero in this approximation order.

Let us now find the second order contributions to $a$ and $\vec w$. The corresponding term in the equation for  $a$ is  $ 3 a^2  \left( \vec \xi^{\dag}, N_{i, j , k, l}  \bfv{Y}_{j} \bfv{\xi}_k \bfv{\xi}_l \right) \vec \xi$. This term has been evaluated numerically and it was confirmed that for the considered spontaneous symmetry breaking bifurcation this term is equal to zero up to the precision of the calculations. At the same time, the component $\vec w$ is not zero in this approximation order and has to be calculated from
$$ \hat L_0 \vec w = -3 a^2  \left( N_{i, j , k, l}  \bfv{Y}_{j} \bfv{\xi}_k \bfv{\xi}_l - ( \vec \xi^{\dag}, N_{i, j , k, l}  \bfv{Y}_{j} \bfv{\xi}_k \bfv{\xi}_l) \vec \xi )\right).$$
The solution of this equation is
\begin{eqnarray}
\vec w = -3 a^2 \sum_i \left( \vec \eta_i^{\dag}  , N_{i, j , k, l}  \bfv{Y}_j \bfv{\xi}_k \bfv{\xi}_l \right) \vec \eta_i / \lambda_i,
\label{appB8}
\end{eqnarray}
where $\vec \eta_i$ and $\vec \eta_i^{\dag}$ are the eigenvectors of $\hat L_0$ and $\hat L_0^{\dag}$ corresponding to nonzero eigenvalues $\lambda_i$ of the operators. Let us mention, that  $\partial_t \vec w$ can be neglected since characteristic time of this motion is much large than the largest inverse absolute value of $\lambda _i$. 

\begin{figure}
\includegraphics[width=\linewidth]{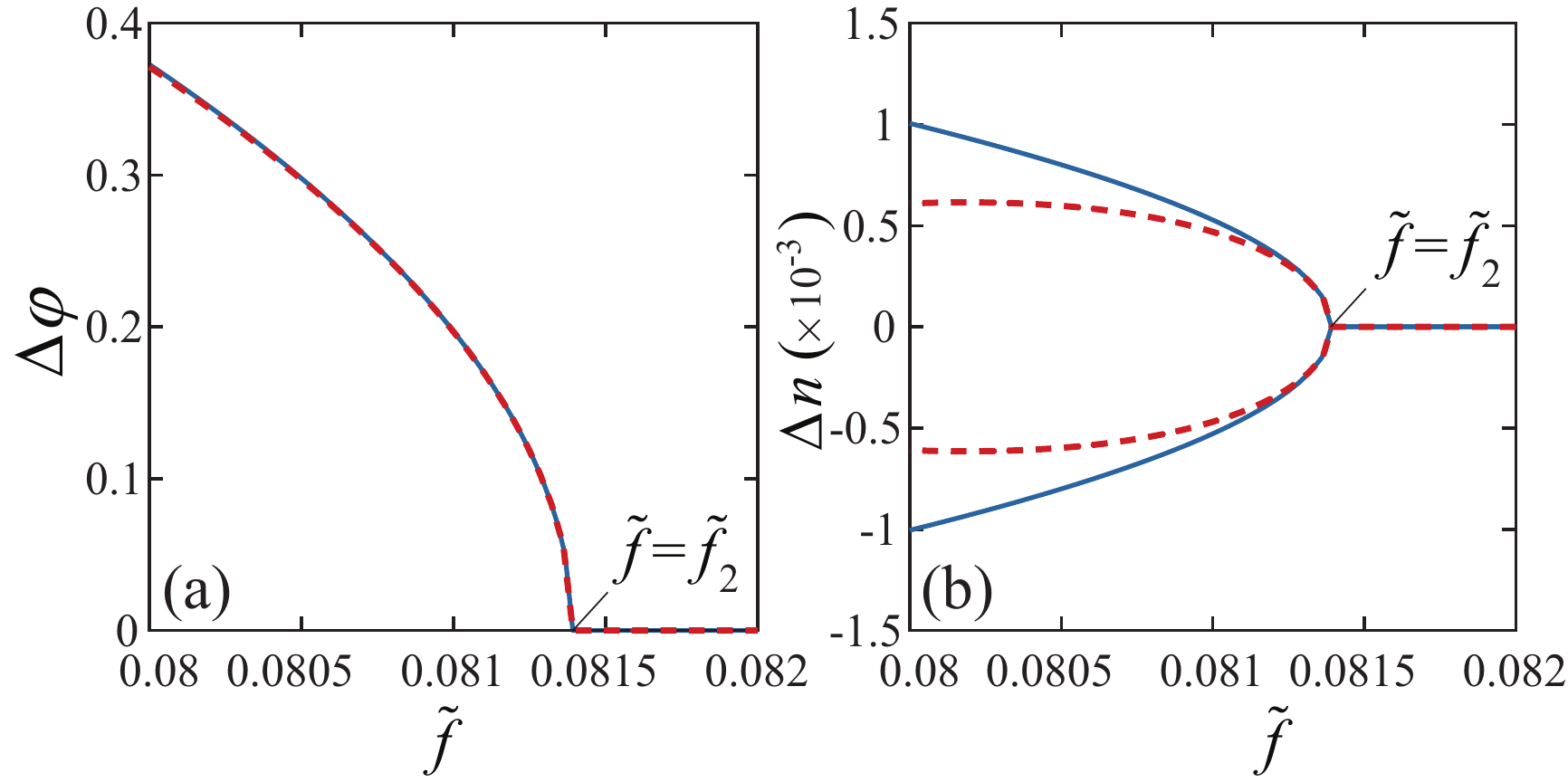}
  \caption{ The phase difference $\Delta\varphi$ (a) and the population imbalance $\Delta n$  in the vicinity of the bifurcation point $\tilde f=\tilde f_2$. The blue curves correspond to the predictions of the two coupled oscillators model \eqref{Eq_toy_2_1} and \eqref{Eq_toy_2_2}. The red curves corresponds to the normal form \eqref{appB9} (equation \eqref{norm_form} in the main text). }
  \label{Fig:NormFormComparison}
\end{figure}

Finally, calculating the third order terms in respect to $a$ in \eqref{appB7} yields:
\begin{equation}
\partial_t a  = \lambda a + \varepsilon a^3,
\label{appB9}
\end{equation}
which matches equation \eqref{norm_form}  from Sec.~\ref{SecIV}.  Here
$$ \varepsilon = \left(\eta_i^{\dag}  , N_{i, j , k, l}  \bfv{\xi}_{j} \bfv{\xi}_k \bfv{\xi}_l \right)  +6 \left(\eta_i^{\dag}  , N_{i, j , k, l}  \bfv{Y}_{j} \bfv{\xi}_k \bfv{w}_l \right).$$

Equation \eqref{appB9} is the normal form of a pitchfork bifurcation. The calculated value of $\epsilon$ for the considered system is negative. Therefore, we conclude that the spontaneous symmetry breaking occurs at the supercritical pitchfork bifurcation leading to the birth of a new stable symmetry broken states. For these states  the complex amplitude of the polariton field  is $a_{\rm s}=\sqrt{-\epsilon\left/\lambda\right.}$. Since $\lambda$ depends linearly on $\tilde f$ we can conclude that the deviation of the symmetry broken state from the symmetric one must show a square root dependency on the detuning from the bifurcation point .

In order to check the validity of the perturbative approach we depict in Fig.~\ref{Fig:NormFormComparison} the phase difference $\Delta\varphi$ and the population imbalance $\Delta n$ calculated with \eqref{Eq_toy_2_1} and \eqref{Eq_toy_2_2} (blue solid curves) and with the normal form equation  \eqref{appB9} (red dashed curves). An excellent coincidence in the vicinity of the bifurcation point justifies the developed perturbation theory.

\section{The analysis of the synchronization scenario in terms of the coupled phase oscillators model}\label{APP_C}
\setcounter{section}{3}

Here we aim to find stationary solutions of equation \eqref{Eq_phase_osc}  and analyze their stability. Only the stable stationary state provides synchronization of the  polariton dyad to the coherent laser light. It is convenient to rewrite \eqref{Eq_phase_osc} in the form
\begin{subequations}
\begin{eqnarray}
\partial_t \phi = \sin\phi \left(f^\prime \cos\theta - 2 \sigma_{\rm d} \cos\phi \right),\label{appc1a}\\
\partial_t \theta = q - \sigma_{\rm J}\cos2\phi + f^\prime \sin\theta \cos\phi,\label{appc1b}
\end{eqnarray}
\end{subequations}
where $\phi=\left(\varphi_2-\varphi_1\right)/2$ and $\theta = \left(\varphi_2+\varphi_1\right)/2-\pi/2$.

The symmetric stationary states implies $\sin \phi=0$. 
Hence according to \eqref{appc1b} there are four independent symmetric solutions:
\begin{subequations}
\begin{eqnarray}
\varphi=0,\ \ \ \theta=-\arcsin{\left(\frac{q-\sigma_{\rm J}}{f^\prime}\right) },\label{appcsym_a}\\
\varphi=0,\ \ \ \theta=\pi+\arcsin{\left(\frac{q-\sigma_{\rm J}}{f^\prime}\right) },\label{appcsym_b}\\
\varphi=\pi,\ \ \ \theta= \arcsin{\left(\frac{q-\sigma_{\rm J}}{f^\prime}\right) },\label{appcsym_c}\\
\varphi=\pi,\ \ \ \theta=\pi - \arcsin{\left(\frac{q-\sigma_{\rm J}}{f^\prime}\right),\label{appcsym_d} }
\end{eqnarray}
\end{subequations}
which exist at $f^\prime >\left|q-\sigma_{\rm J}\right|$, see condition \eqref{eq_symstate_cond} in the main text.

The states with  broken symmetry appear provided that the expression in parenthesis in \eqref{appc1a} vanishes, i.e. at
\begin{equation}\label{appc2}
\cos\phi = \left(f^\prime/2\sigma_{\rm d}\right) \cos\theta.
\end{equation}
Substituting this expression to \eqref{appc1b} one obtains four pairs of the symmetry broken states:
\begin{subequations}
\begin{eqnarray}\label{appc3}
\theta &=& -\frac{1}{2}\left(\chi +\arcsin B\right),\label{appc3a} \\
\theta &=& -\frac{1}{2}\left(\chi +\arcsin B\right)+\pi,\label{appc3b} \\
\theta &=& -\frac{1}{2}\left(\chi -\arcsin B\right)+\frac{\pi}{2},\label{appc3c}\\
\theta &=& -\frac{1}{2}\left(\chi -\arcsin B\right)-\frac{\pi}{2},\label{appc3d}
\end{eqnarray}
\end{subequations}
where $\chi=-\arcsin\left(-\frac{\sigma_{\rm J}/\sigma_{\rm d}}{\sqrt{1+\sigma_{\rm J}^2/\sigma_{\rm d}^2}}\right)$ and $B=\frac{4\sigma_{\rm d}\left(q+\sigma_{\rm J}\left( 1- f^{\prime 2}/4\sigma_{\rm d}^2\right)\right)}{f^{\prime 2}\sqrt{1+\sigma_{\rm J}^2/\sigma_{\rm d}^2}}$. For each value of $\theta$ there are two states with $\phi=\pm \arccos\left(\left({f^\prime}/{2\sigma_{\rm d}}\right)\cos\theta\right)$.

Solutions \eqref{appc3a}-\eqref{appc3d} require $B \leq 1$. This condition is satisfied provided that the  coherent pumping strength exceeds the critical value defined by the following equation:
\begin{equation}\label{appc4}
f^{\prime 4}+8f^{\prime 2} \sigma_{\rm J}\left(q+\sigma_{\rm J}\right) - 16 \sigma_{\rm d}^2 \left( q+\sigma_{\rm J}\right)^2=0,
\end{equation}
which has only one real root given by \eqref{eq_bs_driving} in the main text.

Besides, \eqref{appc2} imposes an additional condition on the existence of the symmetry broken states, namely, $\left({f^{\prime 2}}\left/{4\sigma_{\rm d}^2}\right.\right)\cos^2\theta \leq 1$. Then, expressing $\cos^2\theta$ from \eqref{appc1b}, one obtains
\begin{equation}\label{appc5}
D \pm \sqrt{D^2 - 16 \left( \sigma_{\rm J}+q \right)^2 \left( \sigma_{\rm J}^2+\sigma_{\rm d}^2 \right)}\leq 8\left( \sigma_{\rm J}^2+\sigma_{\rm d}^2 \right),
\end{equation}
where $D=f^{\prime 2} + 4\sigma_{\rm J}(q+\sigma_{\rm J})$. A plus sign in \eqref{appc5} corresponds to the states \eqref{appc3a} and \eqref{appc3b} while minus stands for \eqref{appc3c} and \eqref{appc3d}. In the former case \eqref{appc5} satisfies provided that the conditions
\begin{equation}
f^{\prime}\leq\sqrt{4\sigma_{\rm d}^2 + \left(q-\sigma_{\rm J}\right)^2}\label{appc6a}
\end{equation}
and
\begin{equation}
f^{\prime 2} \leq 8\left( \sigma_{\rm J}^2+\sigma_{\rm d}^2 \right) - 4\sigma_{\rm J} \left(q + \sigma_{\rm J}\right) \label{appc6b}
\end{equation}
are satisfied simultaneously. Equation \eqref{appc6a} defines the position of the pitchfork symmetry breaking bifurcation, see \eqref{eq_f2} in the main text. Expressing $q$ from  \eqref{appc6a} and \eqref{appc6b}, one obtains the range of detuning  \eqref{eq_bs_detuning} where the states \eqref{appc3a} and \eqref{appc3b} exist.

The states \eqref{appc3c} and \eqref{appc3d} exist where either \eqref{appc6b} holds or
\begin{equation}\label{appc8}
f^{\prime} \geq \sqrt{4\sigma_{\rm d}^2 + \left(q-\sigma_{\rm J}\right)^2}.
\end{equation}

The stability analysis of the stationary solutions implies their perturbation in the form $\phi=\phi_0 + \delta\phi \exp\left[\mu^{\rm (L)} t\right]$ and $\theta=\theta_0 + \delta\theta \exp\left[\mu^{\rm (L)} t\right]$, where $\phi_0$ and $\theta_0$ corresponds to the stationary states and $\delta\phi$, $\delta\theta$ are small perturbations. From \eqref{appc1a} and \eqref{appc1b} for the Lyapunov exponent $\mu^{\rm (L)}$ we obtain
\begin{eqnarray}\label{appc9}
&\mu^{\rm (L)}_{1,2}=-\sigma_{\rm d} \cos 2\phi_0 + f^{\prime}\cos \phi_0 \cos\theta_0 \pm \nonumber
\\
 &\pm \left(\sigma_{\rm d}^2 \cos^2 2\phi_0  - \sigma_{\rm J}^2\sin^2 2\phi_0 \vphantom{+\left( f^\prime \sin\phi_0\sin\theta_0 - \sigma_{\rm J}\sin2\phi_0 \right)^2}  \right. \\
\nonumber &\left. +\left( f^\prime \sin\phi_0\sin\theta_0 - \sigma_{\rm J}\sin2\phi_0 \right)^2 \right)^{1/2}.
\end{eqnarray}

For the symmetric solutions:
\begin{equation}\label{appc10}
\mu^{\rm (L)}_{1,2}= -\sigma_{\rm d} \pm  \left|\sigma_{\rm d} \right| + f^{\prime}\cos \phi_0 \cos\theta_0.
\end{equation}
For the states \eqref{appcsym_a} and \eqref{appcsym_d} the last term in \eqref{appc10} is positive. Thus these solutions are unstable, $\mu^{\rm (L)}_{1,2}>0$, at any sign of the dissipative coupling. In contrast, the states \eqref{appcsym_b} and \eqref{appcsym_c} are always stable at $\sigma_{\rm d} >0$. However in the opposite case $\sigma_{\rm d} < 0$, which corresponds to the synchronization of the antisymmetric dyad, they become stable only where the condition \eqref{appc8} is satisfied.

For the symmetry broken solutions \eqref{appc3a}-\eqref{appc3d} the Lyapunov exponents are
\begin{eqnarray}\label{appc11}
&\mu^{\rm (L)}_{1,2}= \sigma_{\rm d} \pm  \nonumber \pm \left(\sigma_{\rm d}^2 \cos^2 2\phi_0  + f^{\prime  2} \sin^2\phi_0\sin^2\theta_0 - \right. \\
& \left. 4\sigma_{\rm J}f^\prime\sin^2\phi_0 \cos\phi_0\sin\theta_0 \right)^{1/2}.
\end{eqnarray}
At the positive dissipative coupling  the stability problem eigenvalue \eqref{appc11} is always positive. Thus, the symmetric dyad can not be synchronized in the symmetry broken state. In the opposite case ($\sigma_{\rm d}<0$), $\mu^{\rm (L)}_2<0$ always, while the first exponent vanishes, $\mu^{\rm (L)}_1=0$, at $\cos \phi_0=1$. The latter condition determines the boundary of the existence domain of  the symmetry broken states, see \eqref{appc5}. The analysis demonstrates that the states \eqref{appc3a} and \eqref{appc3b} remain stable in the domain of their existence, while the states \eqref{appc3c} and \eqref{appc3d}, on the contrary, are always unstable.


%

\end{document}